\documentclass[12pt,preprint2]{aastex}

\newcommand{\Msun}{\hbox{\hbox{M}$_\odot\;$}}

\newcommand{\kms}{\hbox{${\rm km}\:{\rm s}^{-1}\;$}}
\newcommand{\Msuno}{\hbox{\hbox{M}$_\odot$}}

\newcommand{\kmso}{\hbox{${\rm km}\:{\rm s}^{-1}$}}
\newcommand{\teff}{$T_{\rm eff}\;$}  
\newcommand{\teffo}{$T_{\rm eff}$}  
\newcommand{\logg}{$\log\;g\;$}
\newcommand{\loggl}{$\log (g/{\rm cm~s}^2)$}  
\newcommand{\loggo}{$\log\;g$}

\shorttitle{Abundances of Cernis~52}
\shortauthors{Gonz\'alez Hern\'andez et al.}

\begin{document}

\title{The Chemical Composition of Cernis~52 (\mbox{BD+31$^o$ 640})} 

\author{J. I. Gonz\'alez Hern\'andez\altaffilmark{1,2,3}, S.
Iglesias-Groth, R. Rebolo\altaffilmark{4}, D. A.
Garc{\'\i}a-Hern\'andez, and A. Manchado\altaffilmark{4}} 
\affil{Instituto de Astrof{\'\i}sica de Canarias, C/ Via L\'actea s/n,
38200 La Laguna, Spain}
\email{jonay@astrax.fis.ucm.es}

\and

\author{D. L. Lambert}
\affil{The W.J. McDonald Observatory, University of Texas, Austin, TX 78712-1083, USA}          

\altaffiltext{1}{CIFIST Marie Curie Excellence Team}
\altaffiltext{2}{Observatoire de Paris, GEPI, 92195 Meudon Cedex, France}
\altaffiltext{3}{Dpto. de Astrof\'{\i}sica y Ciencias de la Atm\'osfera, Facultad de Ciencias F\'{\i}sicas, Universidad Complutense de Madrid, E-28040 Madrid, Spain}
\altaffiltext{4}{Consejo Superior de Investigaciones Cient\'\i ficas, Spain}

\begin{abstract}
 
We present an abundance analysis of the star Cernis~52 in whose
spectrum we recently reported the napthalene cation in absorption at
6707.4 {\AA}. This star is on a line of sight to the Perseus
molecular complex.
The analysis of high-resolution spectra using a
$\chi2$-minimization procedure and a grid of synthetic spectra
provides the stellar parameters and the abundances of O, Mg, Si,
S, Ca, and Fe. 
The stellar parameters of this star are found to be $T_{\mathrm{eff}} =
8350 \pm 200$ K, \loggl $= 4.2 \pm 0.4$ dex. We derived a metallicity of
$\mathrm{[Fe/H]} = -0.01 \pm 0.15$.  
These stellar parameters are consistent with a star of $\sim 2$~\Msun
in a pre-main-sequence evolutionary stage.
The stellar spectrum is significantly veiled in the spectral range
$\lambda\lambda5150-6730$~{\AA} up to almost 55 per cent of the total
flux at 5150~{\AA} and decreasing towards longer wavelengths.
Using Johnson-Cousins and 2MASS photometric data, we determine a
distance to Cernis~52 of 231$^{+135}_{-85}$~pc considering the error
bars of the stellar parameters. This determination places the star at
a similar distance to the young cluster IC~348. This together with its
radial velocity, $v_r=13.7\pm1$ \kmso, its proper motion and probable
young age support Cernis~52 as a likely member of IC~348. 
We determine a rotational velocity of $v\sin i=65 \pm 5$~\kms for this
star. We confirm that the stellar resonance line of \ion{Li}{1} at
6707.8~{\AA} is unable to fit the broad feature at 6707.4~{\AA}. 
This feature should have a interstellar origin and could possibly
form in the dark cloud L1470 surrounding all the cluster IC~348 at
about the same distance.

\end{abstract} 
  
\section{Introduction}

In a recent study of the diffuse interstellar bands (DIBs) toward a
region of anomalous microwave emission, Iglesias-Groth et al. (2008)
have shown evidence for the presence of the naphthalene cation,
${\rm C}_{10}{\rm H}_8^+$, the simplest Policyclic Aromatic
Hydrocarbon (PAH), in the line of sight towards the moderately
reddened star Cernis~52 (BD+31$^o$ 640, $V=11.4$, $E(B-V)=0.9$, Cernis
1993). The PAHs, including the naphthalene cation, have been also
detected in a cometary dust sample returned to Earth by Stardust and
in interplanetary dust particles, possibly of cometary origin (see the
review by Li 2009).  

Cernis~52 is an early type star, classified as A3V by Cernis (1993),
has coordinates $\alpha = 03^{\rm h}43^{\rm m}00.3^{\rm s}$ and
$\delta = +31^\circ58'26''$ (J2000.0), and is located at an angular
separation of less than one degree from the very young stellar cluster
IC~348 in the Perseus OB2 molecular complex.  
The photometric distance to star Cernis~52 is 240 pc (Cernis 1993),
consistent with a location in the molecular complex OB2 
where the microwave emitting cloud is also most likely located (see
Watson et al. 2005). The uncertainties 
prevent to conclude whether the star is embedded in a  
cloud of this young star forming region or lay behind.

Iglesias-Groth et al. (2008) claimed that the relative strength of
several diffuse interstellar bands detected in the spectrum of
Cernis~52 was consistent with the presence of a molecular cloud 
in the line of sight.  
The anomalous microwave emission detected  
in the direction to Perseus (Watson et al. 2005, 2006) could be
associated to electric dipole radiation of fast spinning hydrogenated
carbon-based molecules in a molecular cloud, a mechanism originally
proposed by Draine and Lazarian (1998). 
It is important to establish whether the cloud responsible 
for the excess color of Cernis~52 hosts the carriers of both, the
diffuse interstellar bands observed in the spectrum of this star and
the anomalous microwave emission detected in its line of sight.

Iglesias-Groth et al. (2008) also noted the near-coincidence in
wavelength between the interstellar  
cation's feature and the stellar Li I resonance line.  
To gain insight into the contribution of the stellar line to the
observed broad feature, we undertook a thorough analysis by
spectrum synthesis of Cernis 52's spectrum with two principal goals -
a general abundance analysis for the star and spectrum synthesis fits to the
spectrum around 6707~{\AA} in order assess the Li I line's
contribution. 

Here we present a precise determination of the stellar parameters
and a detailed chemical composition study of the star Cernis 52. 
This abundance analysis should provide indications of the likely Li
abundance for the star, i.e., if 
the star is not chemically peculiar should have a lithium
abundance no greater than the cosmic Li abundance of 
A(Li)\footnote{A(Li)$=\log [N({\rm Li})/N({\rm H})]+12$}\,$=3.3.$
In this paper we will discuss with special attention the naphthalene
cation's feature at 6707.4~{\AA} in the spectrum of Cernis~52 and 
investigate the relationship of this star with
the Perseus star forming region.

\section{Observations}

We have made use of spectroscopic data obtained 
with the 2dcoud\'{e} cross-dispersed echelle (CS23, Tull et al. 1995) 
at the 2.7 m Harlan J. Smith Telescope at McDonald Observatory (Texas, 
USA). We observed the star Cernis~52 on 26, 27 August 2007, and 1,
2 March 2008 in the spectral range $\lambda\lambda3740-10160$\,{\AA}, with 
resolving power $\lambda/\delta\lambda\sim60,000$ and~$\sim30,000$,
respectively. These two datasets were independently reduced and 
properly combined to produce final spectra of SNR~$\sim$150 and 250, 
respectively, at the position of H$\alpha$ for a dispersion of
0.05\,{\AA}/pixel. 

We also performed spectroscopic observations with the cross-dispersed
echelle spectrograph SARG (Gratton et al. 2001) at the Telescopio
Nazionale Galileo (TNG) at the Observatorio del Roque de los Muchachos
(La Palma, Canary Islands, Spain). We observed the target star on 4
November 2006 in the spectral ranges $\lambda\lambda4600-7975$ and
$\lambda\lambda5600-10100$\,{\AA}, 
with a spectral resolution of $\lambda/\delta\lambda\sim57,000$ and
a final SNR~$\sim 100$ at H$\alpha$ for a dispersion of
0.034\,{\AA}/pixel.

Early-type rapidly rotating hot stars were also 
observed for correction of possible telluric absorptions in
all campaigns. All the spectra were reduced within IRAF, wavelength
calibrated and co-added after correction for telluric lines.

\section{Rotational and radial velocity\label{secrot}}

Firstly, each of the three final spectra was cross-correlated
(within the package MOLLY) with a synthetic spectrum properly
broadened with a $v\sin i=80$~\kmso. Thus, we corrected these spectra
for their radial velocities and derived a rotational velocity of
$v\sin i=65 \pm 5$~\kmso by comparing broadened and veiled versions 
of a A3V synthetic spectrum in steps of 5~\kmso, adopting an
spherical rotational profile with linearized  
limb-darkening $\epsilon = 0.46$, appropriate to the spectral type of
the star (Al-Naimiy 1978).
 
Finally, we estimate a radial velocity of Cernis~52 at $v_r=+13.7 \pm 
1$~\kmso. The cross-correlation was performed in
the spectral range $\lambda\lambda5140-6700$~{\AA} and masking the broad H$\alpha$
profile and the interstellar features at 5889-95~{\AA} and
6280~{\AA}. 
This determination is consistent with the mean
radial velocity of the young cluster IC~348, which has been recently
estimated at $13.5\pm1.2$~\kms from 30 near solar-mass cluster
members (Dahm 2008). However, Nordhagen et al. (2006) derived a mean
cluster velocity of $15.9\pm0.8$~\kmso, also from 30 cluster members.

We also determine a radial velocity of $14.2\pm1.4$~\kms from the
narrow atomic interstellar lines \ion{Na}{1}~D 5889-95~{\AA},
\ion{Li}{1} 6708~{\AA} and \ion{K}{1} 7698~{\AA}.
The uncertainty on the velocity comes from the dispersion of these
velocity measurements.

These determinations share the velocity of the most luminous cluster
member HD281159 ($\sim14$~\kmso, Evans 1967) and an expanding
supernova shell in which the young cluster IC~348 
is embedded ($\sim12-15$~\kmso, Snow et al. 1994). 
In addition, these atomic interstellar lines have 
a velocity consistent with that of the interstellar hyperfine
transitions of the molecule ${\rm NH}_3$ measured in radio wavelegths
(Rosolowsky et al. 2008): 
the LSR velocities are $7.5\pm1.4$~\kms from the ISM lines
and $6.1 - 8.4$~\kms from the radio transitions.

\section{Companion star\label{seccont}}

\begin{figure}[!ht]
\centering
\includegraphics[width=7.5cm,angle=0]{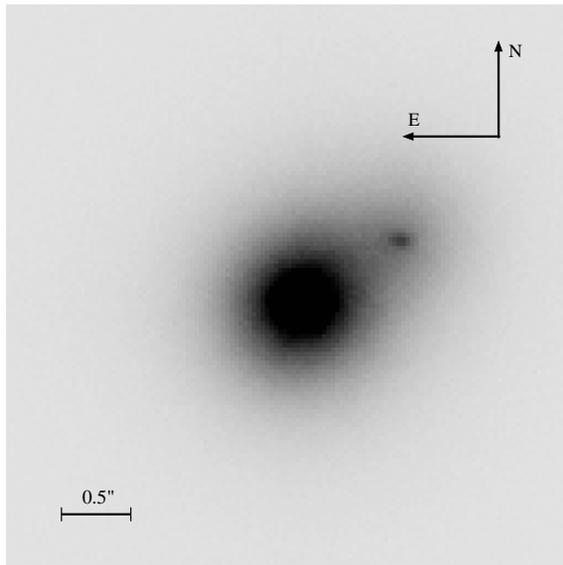}
\caption{FastCam image in the $R$ band of Cernis~[93]52. 
The displayed field size is $\sim 3.5\arcsec\times3.5\arcsec$. 
There exists a companion star at a distance of
$818\pm7$~mas in the direction North-West of Cernis~52.}  
\label{c52r}
\end{figure}

Using the camera FastCam (Oscoz et al. 2008) attached to the Nordic
Optical Telescope (NOT) at the Observatorio del Roque de los
Muchachos, we obtained images in $V$, $R$ and $I$
bands of Cernis~52 on 25 October 2008. We found a companion star at a
distance of $818\pm7$~mas. This camera is able to obtain and
record series of thousands of short exposures, and with a specially
designed software can reach angular resolutions of 
roughly $0.1\arcsec$. 
In Fig.~\ref{c52r} we display the resulting image in the $R$ band. 
We have zoomed in the image to show the two stars 
clearly resolved. The camera has 512x512 pixels and a pixel
size of $31.01\pm0.01$~mas. The companion star is located in the
North-West of Cernis~52, with a position angle North-East
$\theta=325.9^\circ$. 

The companion star is roughly $1.7\pm0.2$
mag fainter than Cernis~52 in the $R$ band. In our spectroscopic
observations with the instrument CS23, obtained prior to the 
discovery of this companion star, the slit widths were $1.2\arcsec$
and $2.4\arcsec$ for the 2007 and 2008 campaigns,
respectively. The slit length was $8.2\arcsec$ in both campaigns. 
Since the slit width was quite large and the seeing during the
observations was typically $\simeq 2-2.5\arcsec$, most of the
light from the companion star must have entered in the slit and
contaminated the spectrum of Cernis~52. 
Assuming that the primary is A3V and that both stars are a physical
pair we can estimate the spectral type of the companion from the
observed magnitude difference in the R-band. We adopt a
\teffo\,$=8500$~K and \loggo\,$=4.0$ for the primary. Thus, we find a
probable \teffo\,$=5800$~K for the companion from suitable
bolometric corrections (Bessell et al.  
1998) and assuming that both stars share the same color excess.
Thus, we search for evidence of the spectrum of a
solar-type companion in the 2007 and 2008 final spectra of Cernis~52. 
We cross-correlated these spectra with the spectrum of the Sun (Kurucz
et al. 1984), properly broadened with different rotational velocities
and in a radial velocity range from -150 to 150 \kmso,
but we did not find any clear signature of this star in the
cross-correlation function.  

It is likely that the companion star have slightly contaminated
the spectrum of Cernis~52. 
From the magnitude difference in the $R$ band and taking
into account that the companion star was partially out from the slit,
we estimate the flux ratio of both stars in our spectrum at $F_{\rm
sec}/F_{\rm C52}=0.11\pm0.05$ in the wavelength region corresponding
to the $R$ band. However, as we will see this is not
sufficient to explain the level of veiling found in the analysis of
the photospheric features of Cernis~52. There must be another reason
for the apparent weakness of the stellar features. 

\begin{figure}[!ht]
\centering
\includegraphics[width=7.5cm,angle=0]{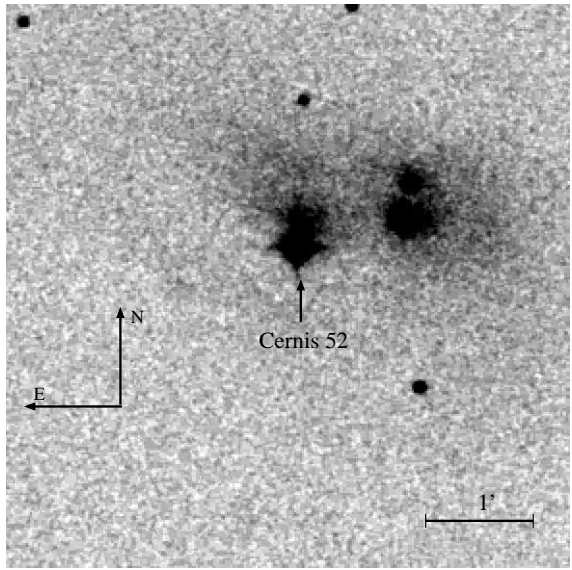}
\caption{DDS-2 image in the $R$ band of Cernis~[93]52. 
The displayed field size is $\sim 5\arcmin\times5\arcmin$.}  
\label{c52dss2r}
\end{figure}

In Fig.~\ref{c52dss2r} we display an optical image from the Palomar
Digital Sky Survey\footnote{http://archive.eso.org/dss/dss.} of
Cernis~52 with a field size of $5\arcmin\times5\arcmin$. In this DSS-2 
red image one can see an entended emission of interstellar gas
surrounding the star which may be the reason why the spectrum of
Cernis~52 is veiled. Our spectra were not background subtracted, since
the star is very bright and the raw images do not show any clear
sky line. 
However, in the raw images there is neither any indication of
the presence of this nebulosity, appart from the high number of
counts of the sky, and any signal must have been subtracted when
correcting for the scattered light. In Fig.~\ref{c52dss2r} we see 
that the interstellar cloud is more prominent towards the North 
of the star, $\sim0.5\arcmin$ from the star center. However, we do
not see any asymmetry in the signal of the sky spectrum which may
indicate that the slit was not positioned South$-$North. The
interstellar cloud may have contaminated our spectra but we suspect
that most of the light coming from the cloud must have been
subtracted during the data reduction. 
A discussion of the interstellar emission in the proximity of
Cernis~52, using Spitzer data, is deferred to a forthcoming paper
(S. Iglesias-Groth et al. 2009, in preparation).

In the next section, we will concentrate on the determination of the
stellar parameters of Cernis~52, taking into account any possible
source of veiling in the spectrum, whose flux we will denote as
$F_{\rm veil}$. 

\section{Stellar Parameters}

\begin{table}
\centering
\caption[]{Ranges and steps of model parameters\label{tblpar}}
\begin{tabular}{lcc}
\noalign{\smallskip}
\noalign{\smallskip}
\noalign{\smallskip}
\hline
\hline
\noalign{\smallskip}
Parameter & Range & Step\\
\noalign{\smallskip}
\hline
\noalign{\smallskip}
$T_{\mathrm{eff}}$ (K)  & $7850 \rightarrow 8750$ & 100\\ 
$\log (g/{\rm cm~s}^2)$  & $3. \rightarrow 5$  & 0.1\\ 
$\mathrm{[Fe/H]}$ & $-0.5 \rightarrow 0.5$  & 0.05\\ 
$f_{6562}$ &  $0 \rightarrow 1.0$  & 0.05\\ 
$m_0$ ({\AA}$^{-1}$) & $0 \rightarrow -0.001$ & -0.0001\\ 
\noalign{\smallskip}
\hline	     
\hline
\end{tabular}
\end{table}

We have inspected the spectrum of Cernis~52 to search for 
atomic stellar lines appropriate to provide accurate 
element abundances. We realized that the stellar lines were
significantly veiled even at a level of $\sim 50$ per cent 
at 5180~{\AA} where the \ion{Mg}{1}b triplet is located.
Thus, the Fe abundances obtained from the lines \ion{Fe}{1} 5214 {\AA}
and \ion{Fe}{2} 5273 {\AA} if we do not include any veiling are
[Fe/H]\,$=-0.55$ and -0.8, respectively. This metallicity seems to be
too low for the location of this star in a star-forming region. 
In addition, the \ion{Mg}{1}b 5172 {\AA}
would require an abundance of [Mg/Fe]$\sim-1.0$. These results are
achieved for all reasonable values of the atmospheric parameters,
\teff and \loggo.
The analysis of this spectrum requires a tool able to take 
into account the effect of any possible veiling when computing
atmospheric abundances. 
To simplify, we assumed the veiling to be a linear function of
wavelength, and thus defined 
by two parameters, its value at 6562\,{\AA}, $f_{6562}=F^{6562}_{\rm
veil}/F^{6562}_{\rm C52}$, and the slope, $m_0$. Note that the total
flux is defined as $F_{\rm total} = F_{\rm veil}+F_{\rm C52}$, where
$F_{\rm veil}$ and $F_{\rm C52}$ are the flux contributions of the
source of veiling and the continuum of Cernis~52, respectively.
We decided to split the determination of the
stellar and veiling parameters in two steps: 1) using the H$\alpha$
profile to derive the stellar temperature and the veiling at 6562\,{\AA};
and 2) using \ion{Fe}{1} and \ion{Fe}{2} features to derive the surface
gravity, the slope of the veiling and the metallicity of Cernis~52.

\begin{figure}[!ht]
\centering
\includegraphics[width=8.4cm,angle=0]{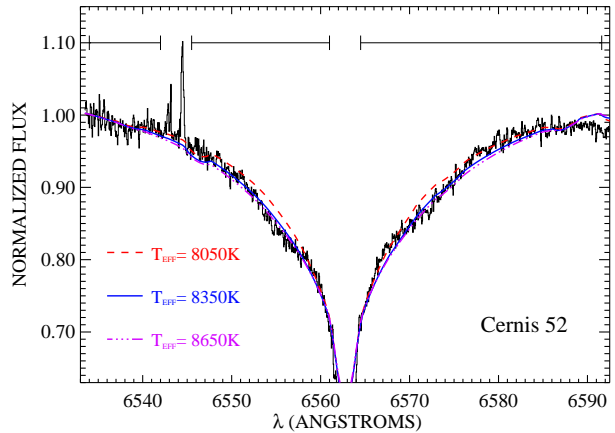}
\caption{Synthetic H$\alpha$ profiles for three effective
temperatures compared to the observed spectrum of \mbox{Cernis~52}
(SNR $\sim 150$) and normalized to the level of the observed spectrum
of \mbox{Cernis~52} at 6540\,{\AA}. We also depict the fitting
regions at the top.} 
\label{halpha}
\end{figure}

\subsection{H$\alpha$ Profile\label{sechalpha}}

The wings of H$\alpha$ are a very good temperature indicator 
(e.g., Barklem et al. 2002). Adopting the theory of Ali \& Griem (1965,
1966) for resonance broadening and Griem (1960) for Stark broadening,
we computed H$\alpha$ profiles for several effective temperatures,
using the code SYNTHE (Kurucz 2005; Sbordone 2005). For
further details on the computations of hydrogen lines in SYNTHE,
see Castelli \& Kurucz (2001) and Cowley \& Castelli (2002).
 
In this section, we will use the spectrum of Cernis~52 taken in 2007
since the 2008 spectrum had a different wavelength setting and the
H$\alpha$ line appears in an interorder region. In Fig.~\ref{halpha} we
compare the synthetic H$\alpha$ profiles with the observed profile for
several temperatures. To evaluate the goodness of fit, we employ a
reduced $\chi2$ statistic, 
\begin{equation}
\chi_\nu^2 = \frac{1}{N-M}\sum^N_{i=1}\left(\frac{f_i-F_i}{\sigma_i}\right)2,
\end{equation}
where $N$ is the number of wavelength points, $M$ is the number of free
parameters (here two, \teff and $f_{6562}$),
$f_i$ is the synthetic normalized flux, $F_i$ is the observed 
normalized flux, and $\sigma_i=1/{\rm SNR}$. The SNR was estimated as
a constant average value in continuum regions close to the observed
H$\alpha$ profile. The fitting regions are indicated in
Fig.~\ref{halpha}, which contains all of the spectral regions close
the center of the H$\alpha$ profile where there are no stellar lines
and where the normalized flux is greater than $\sim0.7$. This figure
also shows that the sensitivity of this method to the \teff decreases
toward higher effective temperatures. 

\begin{figure}[ht!]
\centering
\includegraphics[width=8.4cm,angle=0]{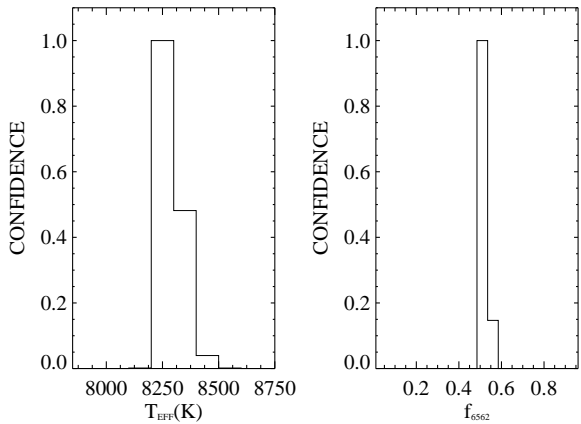}
\caption{Normalized distributions obtained for the effective
temperature, \teffo, measured from the H$\alpha$ profile and the
veiling at 6562\,{\AA}, $f_{\rm 6562}$, using Monte Carlo simulations.
The total number of simulations was 1000.} 
\label{parhalpha}
\end{figure}

\begin{figure}[ht!]
\centering
\includegraphics[width=8.4cm,angle=0]{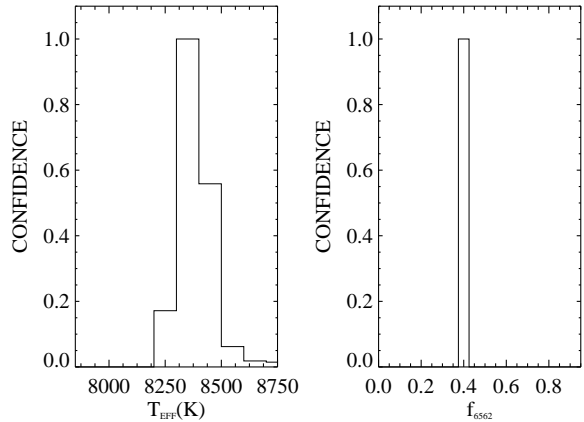}
\caption{Same as Fig.~\ref{parhalpha}, but including the spectrum of
the Sun with a contribution of 10\% of the stellar flux.} 
\label{parhalphasun}
\end{figure}

In order to estimate the uncertainties on temperature and veiling, 
we built 1000 realizations of the observed spectrum taking into
account the SNR. We used a boostrap Monte Carlo method to define the
1-$\sigma$ confidence regions and found as most likely values 
$T_{\mathrm{eff}} = 8300 \pm 100$ K and $f_{6562} = 0.50\pm0.05$. The
corresponding histograms are displayed in Fig.~\ref{parhalpha}. The
\teff and $f_{6562}$ were varied within the ranges given in
Table~\ref{tblpar}. To check the influence of the spectrum of the
companion star we perform another simulation by introducing the
H$\alpha$ profile of a pre-main sequence solar metallicity star with
\teff$= 5800$ K and \logg$=3.8$, and assuming that this spectrum was
contributing with the 10\% of the stellar flux. The adopted surface
gravity for the companion star coincides with the expected value for a
pre-main sequence star of this temperature (see Section~\ref{secabun}). 
We use a rotational velocity of $v\sin i=100$~\kms and a limb-darkening coefficient
$\epsilon = 0.6$, because the spectrum of Cernis~52 does not show any
stellar narrow absorption and also to be consistent with the synthetic
spectra computed for Section~\ref{secabun}. However, we check that
using lower rotational velocities does not affect the derived \teffo. 
The result is shown in Fig.~\ref{parhalphasun}. The distribution of
temperatures slightly changes but our adopted effective temperature
remains almost the same, $T_{\mathrm{eff}} = 8350 \pm 100$ K. In
contrast, the required veiling factor changes, being in this case
$f_{6562} = 0.40\pm0.05$. 

\subsection{Ionization Equilibrium of Iron\label{seciron}} 

Once we have determinated the veiling at 6562\,{\AA} and the effective
temperature we try to derive the surface gravity and metallicity using
\ion{Fe}{1} and \ion{Fe}{2} features of the spectrum of Cernis~52. In
this case, we will neglect the presence of a companion spectrum since
it hardly affects the spectrum of Cernis~52.
We use a technique which compares a grid of synthetic spectra with the
observed spectrum, via a $\chi2$-minimization procedure. In this case,
we will use the spectrum of Cernis~52 taken in 2008, since it has
higher SNR. This procedure takes into account the effect that veiling
causes on the stellar lines and it has been applied in X-ray binaries
to derive the stellar parameters of the companion stars in these
systems (A0620--00, Gonz\'alez Hern\'andez et al. 2004; Centaurus X-4,
Gonz\'alez Hern\'andez et al. 2005; XTE J1118+480, Gonz\'alez
Hern\'andez et al. 2006, 2008b; and Nova Scorpii 1994, Gonz\'alez
Hern\'andez et al. 2008a). However, we 
needed to slightly modify this program in order to allow a fixed
veiling at a given wavelength. In the previous version of this
program, the slope of the veiling was defined using a given veiling at
4500\,{\AA}. We typically produce 10 different values for the slope,
whose maximum value is zero, i.e. a constant veiling, and the minimum
value provides a veiling of $f_{6562}=0.50$ and $f_{5000}=2$
(see Table~\ref{tblpar}).

We already know the veiling at 6562\,{\AA}, but we need also to
find the slope of the veiling in order to properly model the observed 
features.
We selected 10 spectral features of iron in the spectral range
$\lambda\lambda5270-6400$\,{\AA}, containing in total 55 lines of
\ion{Fe}{1} and 25 lines of \ion{Fe}{2} with excitation potentials
between 0.5 and 10.5 eV. We compute
synthetic spectra with the local thermodynamic equilibrium (LTE) code
MOOG (Sneden 1973), adopting the atomic line data from the Vienna
Atomic Line Database (VALD, Piskunov et al. 1995) and using a grid of
LTE model atmospheres (Kurucz 1993). 
The oscillator strengths of the iron lines as well as
other element lines used in this work were adjusted until 
reproducing both the solar atlas of Kurucz et al. (1984) with solar
abundances (Grevesse et al.\ 1996) and the spectrum of Procyon with
its derived abundances (Andrievsky et al.\ 1995).
We generated a grid of synthetic spectra for
the \ion{Fe}{1} and \ion{Fe}{2} features varying as free parameters,
the star effective temperature ($T_{\mathrm{eff}}$), the surface
gravity ($\log g$), the metallicity ([Fe/H]), and the veiling,
charaterized by the $f_{6562}$ and $m_0$. 
The microturbulence,
$\xi=2$\,\kmso, was fixed in each atmospheric model according to the
spectral type of Cernis~52 (see e.g. Fossati et al. 2008). 
We choose an error of 0.5\,\kms for the microturbulent velocity.

\begin{figure*}[ht!]
\centering
\includegraphics[width=14.cm,angle=0]{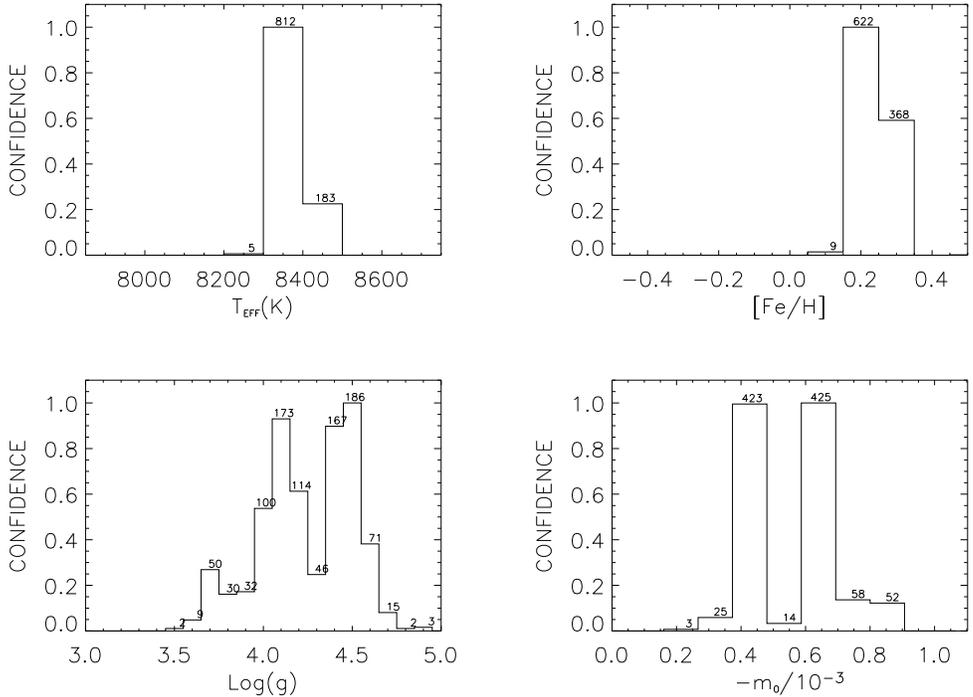}
\caption{Normalized distributions obtained for the effective
temperature, \teffo, surface gravity, \loggo, metallicity, [Fe/H], and
veiling slope, $m_0$, using Monte Carlo simulations. The labels at the
top of each bin indicate the number of simulations consistent with the
bin value. The total number of simulations was 1000.} 
\label{parall}
\end{figure*}

We fixed the veiling parameter at $f_{6562}=0.50$ and allowed the
effective temperature to be in the range $8250 <$\,\teffo\,$<8450$. 
The other three free parameters were varied using the ranges and steps
given in Table~\ref{tblpar}. We use a bootstrap Monte-Carlo method to
derive the resulting histograms of these parameters shown in
Fig.~\ref{parall}, corresponding again to 1000 realizations.  
We found strong evidence of high veiling in the spectral region under
analysis (5270--6400\,{\AA}).
This veiling increases toward shorter wavelengths, being
as high as $f_{5180}\sim1.2$ at the \ion{Mg}{1}b
triplet 5167--83{\AA}. This translates into a veiling of roughly 54\% when
computing the ratio of the flux $F^{5180}_{\rm veil}$ to the total
flux, $F^{5180}_{\rm total}$. Therefore, at 5180~{\AA} the star
contributes almost with half of the total flux. 
By inspecting the histograms displayed in Fig.~\ref{parall}, we
decided to choose as most likely values:
$T_{\mathrm{eff}} = 8350 \pm 200$ K, $\log (g/{\rm cm~s}^2) = 4.2 \pm
0.4$, $\mathrm{[Fe/H]} = 0.20 \pm 0.15$, $f_{6562} = 0.50\pm0.10$, and
$m_0 = -0.00055\pm0.00010$. These stellar parameters are consistent 
with the spectral type already reported by Cernis~(1993).

The surface gravity is the stellar parameter that is usually
more difficult to determine when dealing with fast rotating and veiled
stars (see e.g. Gonz\'alez Hern\'andez et al. 2005, 2008b).
We warn the reader that the error bar given for the surface gravity
comes from the range of values that shows the histogram of this
parameter in Fig~\ref{parall}.

\begin{deluxetable}{lrrrrrrrrrrr}
\tabletypesize{\scriptsize}
\tablecaption{Chemical Abundances of Cernis~52}
\tablewidth{0pt}
\tablehead{Species & $\log
\epsilon(\mathrm{X})_{\odot}$\tablenotemark{a} & $[{\rm X}/{\rm H}]$ &
$[{\rm X}/{\rm Fe}]$ & ${\sigma}$ & $\Delta_{\sigma}$ & $\Delta_{T_{\rm
eff}}$ & $\Delta_{\log g}$ & $\Delta_{\rm veil}$ & 
$\Delta_{\xi}$ & $\Delta \mathrm{[X/H]}$ & $N$\tablenotemark{b}} 
\startdata
 \ion{O}{1} &  8.74 &  0.54 &  0.55 &  0.28 &  0.20 &  0.00 &  0.03 &  0.15 & -0.10 &  0.27 &    2 \\ 
\ion{Mg}{1} &  7.58 &  0.35 &  0.36 &  0.25 &  0.25 &  0.20 & -0.05 &  0.15 & -0.30 &  0.47 & 5173 \\ 
\ion{Si}{2} &  7.55 & -0.10 & -0.09 &  0.28 &  0.20 &  0.00 &  0.03 &  0.10 & -0.15 &  0.27 &    2 \\ 
 \ion{S}{1} &  7.33 & -0.19 & -0.18 &  0.04 &  0.02 &  0.10 &  0.00 &  0.05 &  0.00 &  0.11 &    2 \\ 
\ion{Ca}{1} &  6.36 & -0.02 & -0.01 &  0.22 &  0.08 &  0.17 & -0.02 &  0.07 & -0.05 &  0.21 &    7 \\ 
\ion{Ca}{2} &  6.36 &  0.10 &  0.11 &  0.25 &  0.25 &  0.10 &  0.10 &  0.05 &  0.00 &  0.29 & 5307 \\ 
\ion{Fe}{1} &  7.50 &  0.03 &  0.04 &  0.19 &  0.05 &  0.13 &  0.01 &  0.05 & -0.03 &  0.15 &   13 \\ 
\ion{Fe}{2} &  7.50 & -0.06 & -0.05 &  0.08 &  0.03 &  0.07 &  0.04 &  0.04 & -0.06 &  0.11 &    8 \\ 
\enddata

\tablecomments{Chemical abundances of Cernis~52 and uncertainties
produced for $\Delta_{T_{\rm eff}} = +200$\,K, $\Delta_{\log g} = +0.4$
dex, $\Delta_{\rm veil} = +0.1$, and $\Delta_{\xi} = +0.5~$\kmso. 
All the element abundances were determined by fitting the
observed spectra with synthetic spectra computed with the LTE code
MOOG.}

\tablenotetext{a}{The solar element abundances were adopted from
Grevesse et al. (1996) and Ecuvillon et al. (2006)} 

\tablenotetext{b}{Number of spectral features of this atomic specie in
the star, or if there is only one, its wavelength.}

\label{tblabu}      
\end{deluxetable}

\begin{figure*}[!ht]
\centering
\includegraphics[width=12cm,angle=90]{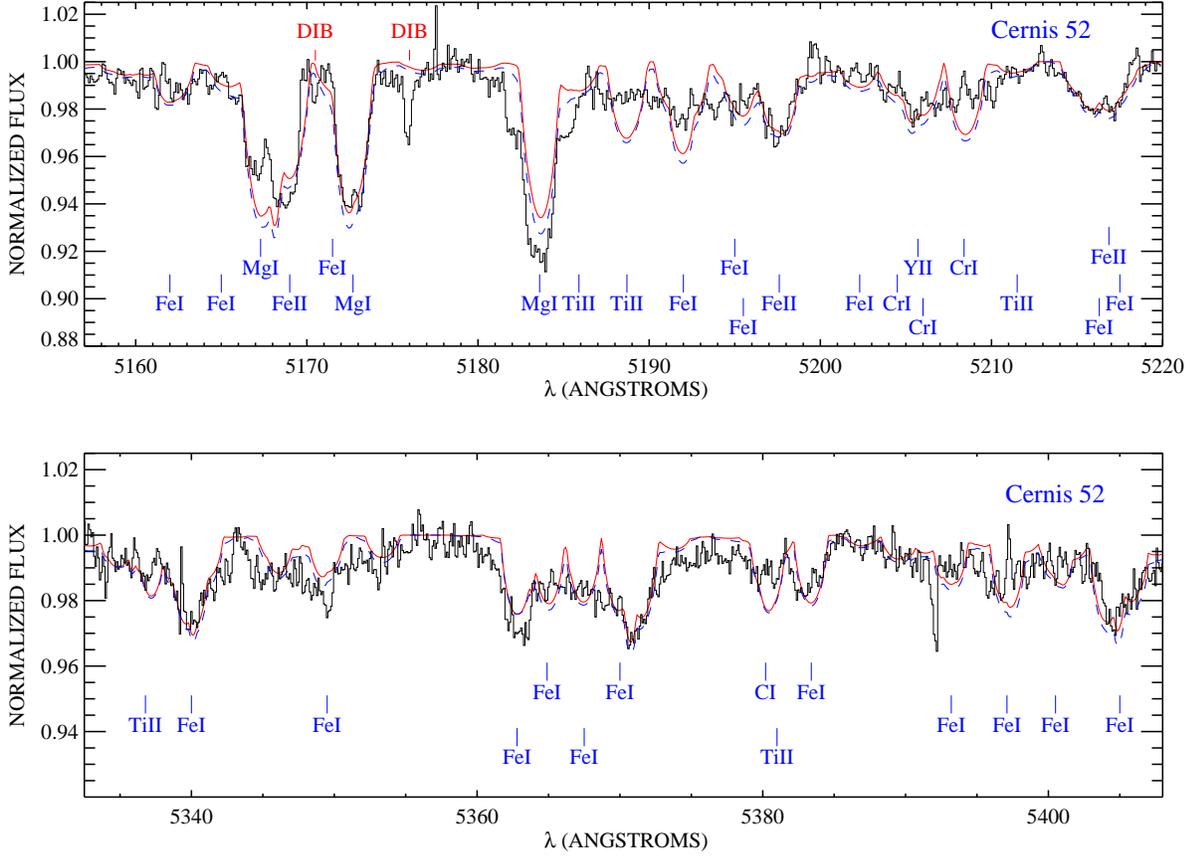}
\caption{Best synthetic spectral fits (solid lines) to the observed
spectrum of \mbox{Cernis~52} for two spectral regions. We also
display the combined spectrum (dashed line) including the spectrum of
Cernis~52 contributing with 92\% (top panel) and 91\% (bottom panel)
of the stellar flux and the spectrum of the companion star contributing
at 8 and 9\%, respectively. Note that the veiling factors
$f_{5180}=1.25$ and $f_{5370}=1.15$ have been adopted. The label
``DIB'' refers to diffuse interstellar bands, whereas other labels
show stellar absorption lines of Cernis~52. 
}
\label{figmg}
\end{figure*}

\begin{figure*}[!ht]
\centering
\includegraphics[width=12cm,angle=90]{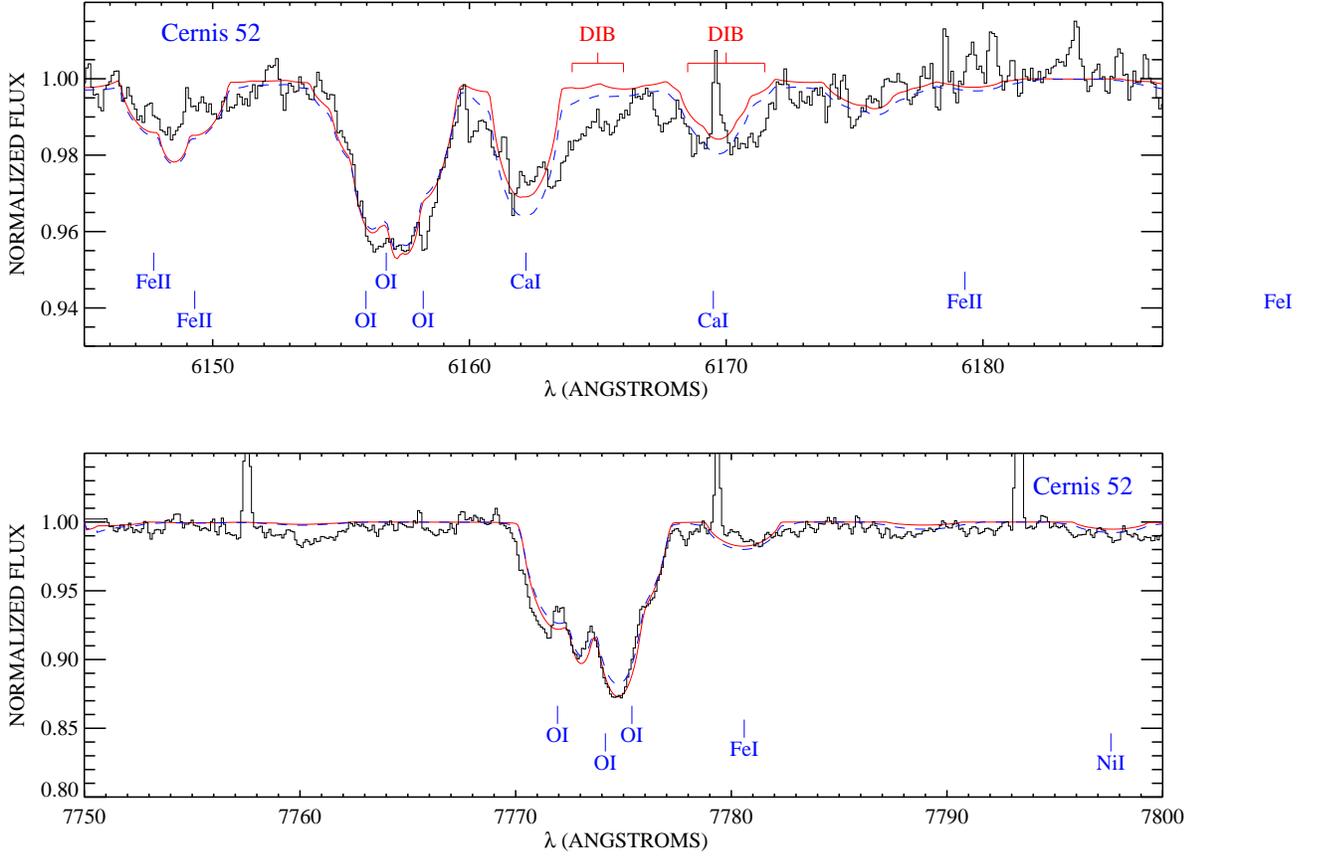}
\caption{Same as Fig.~\ref{figmg}, but for other spectral
regions. In this case, the best fit abundance of each feature has been
used, being [O/H]\,$=0.74$ in the upper panel and [O/H]\,$=0.34$ in the
lower panel. We also display the combined spectrum (dashed line)
including the spectrum of Cernis~52 contributing with 90\% (top panel)
and 87\% (bottom panel) of the stellar flux and the spectrum of the
companion star contributing at 10 and 13\%, respectively. Note that
the veiling factors $f_{6157}=0.75$ and $f_{7773}=0$ have been
adopted.}   
\label{figo}
\end{figure*}

\begin{figure*}[!ht]
\centering
\includegraphics[width=12cm,angle=90]{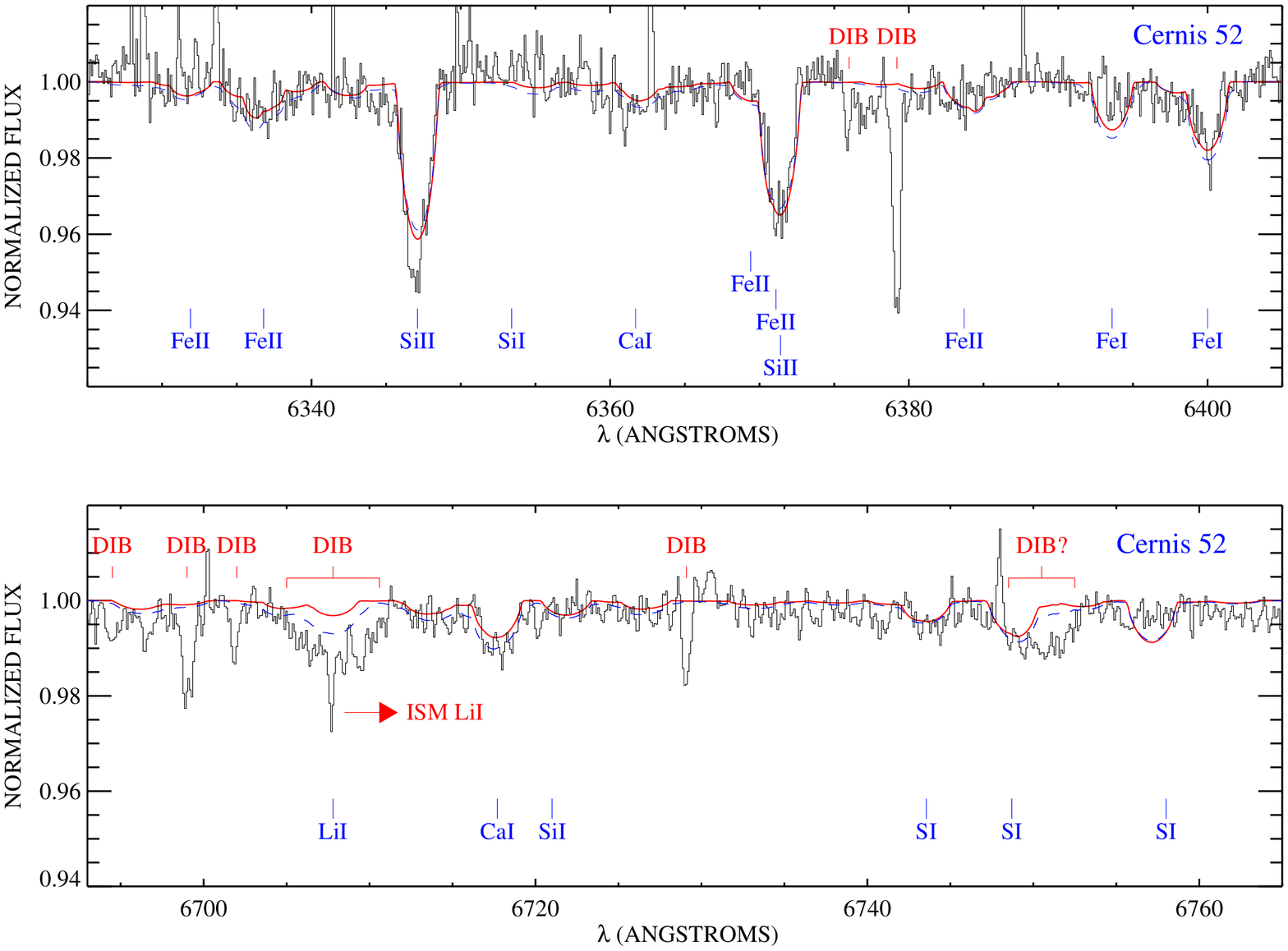}
\caption{Same as Fig.~\ref{figmg}, but for other spectral
regions. We also display the combined spectrum (dashed line) including
the spectrum of Cernis~52 contributing with 90\% (both panels) of the
stellar flux and the spectrum of the companion star contributing at
10\%, respectively. In the bottom panel, the solid line show a
synthetic spectrum of Cernis~52 with A(Li)\,$=3.3$, whereas the dashed
line shows a combined spectrum in which both Cernis~52 and the
companion star have A(Li)\,$=3.3$. 
Note that the veiling factors $f_{6360}=0.60$ and $f_{6720}=0.40$ have
been adopted. The label ``ISM LiI'' gives the location of the atomic
Li interstellar line. 
}
\label{figsi}
\end{figure*}

\section{Photospheric abundances\label{secabun}}

Using the derived stellar and veiling parameters, we firstly determined 
the abundance of Fe from each individual feature of \ion{Fe}{1} and
\ion{Fe}{2} in the observed spectrum (see Table~\ref{tblabu}). We made
use of a $\chi2$ procedure to find the best fit to the 
observed features as in Gonz\'alez Hern\'andez et al. (2004, 2008a). 
In Fig.~\ref{figmg} we show some of the spectral regions analysed to
obtain the Fe abundance. To determine the abundances of other
elements, we adopted the average abundance of the Fe abundances
extracted from \ion{Fe}{1} and \ion{Fe}{2} features, ${\rm
[Fe/H]}=-0.01\pm0.15$, for the metallicity of the model.  

In Table~\ref{tblabu} we provide the average abundance of each element
together with the errors. The errors in the element abundances show
their sensitivity to the uncertainties in the effective temperature
($\Delta_{T_{\mathrm{eff}}}$), surface gravity ($\Delta_{\log g}$),
microturbulence ($\Delta_{\xi}$), veiling ($\Delta_{\rm veil}$) 
and the dispersion of the abundance measurements from different
spectral features ($\Delta_{\sigma}$).  
The errors $\Delta_{\sigma}$ were estimated as $\Delta_{\sigma}
=\sigma/\sqrt{N}$, where $\sigma$ is the standard deviation of 
the $N$ measurements. For those element species for which only one
spectral feature was available we adopt $\sigma=0.25$.
The total error in the abundance of each element
is determined by adding in quadrature all these individual errors.

In Figs.~\ref{figmg},~\ref{figo} and~\ref{figsi} we display several
spectral regions of the observed spectrum of \mbox{Cernis~52} in
comparison with synthetic spectra computed using the derived abundances,
except for oxygen, for which the best fit abundance was used for each
feature. In these figures, we also show the combined synthetic
spectrum of Cernis~52 and the companion star in which the companion
star only contributes with $\sim 10$\% of the stellar flux, and
using the same veiling factors as for the single synthetic 
spectrum of Cernis 52. The
stellar parameters, metallicity and radial velocity of the companion
star were already specified in Section~\ref{sechalpha}. 
We assumed that the radial velocity of the companion star is the same
as for Cernis~52. For most of the features, the presence of the
companion do not change significantly the line profiles, except for the
\ion{Mg}{1}b 5167--83{\AA} lines and the \ion{Ca}{1} features. We note
that we have not taken into account the presence of the companion star
in the abundance determination, since this effect would be
negligible and in any case, the abundances would remain within the
error bars given in Table~\ref{tblabu}. 

\begin{deluxetable}{lrrrrrr}
\tabletypesize{\scriptsize}
\tablecaption{Distance to Cernis~52\label{tbldist}}
\tablewidth{0pt}
\tablehead{\colhead{Parameter} & \colhead{$E(B-V)\tablenotemark{a}=0.9$}  &
\colhead{\teff$\tablenotemark{b}=8550$\,K}  & \colhead{\logg$\tablenotemark{b}=3.8$}  &
\colhead{\logg$\tablenotemark{b}=4.6$}  & \colhead{$E(B-V)\tablenotemark{b}=0.8$}  &
\colhead{$M_{\star}\tablenotemark{b}=2.2~\Msuno$}} 
\startdata
$d_{m_V}\tablenotemark{c}$ (pc) & $237\pm 9$ & $247\pm 9$ & $378\pm14$ & $149\pm 9$ & $273\pm 5$ & $249\pm10$ \\
$d_{m_R}$ (pc)                  & $229\pm10$ & $237\pm10$ & $363\pm16$ & $144\pm11$ & $256\pm 6$ & $240\pm11$ \\
$d_{m_I}$ (pc)                  & $247\pm10$ & $253\pm11$ & $390\pm17$ & $157\pm11$ & $268\pm 6$ & $259\pm11$ \\
$d_{m_J}$ (pc)                  & $233\pm 2$ & $237\pm 2$ & $369\pm 3$ & $147\pm 2$ & $242\pm 1$ & $245\pm 2$ \\
$d_{m_H}$ (pc)                  & $232\pm 2$ & $235\pm 2$ & $367\pm 4$ & $147\pm 3$ & $238\pm 1$ & $243\pm 2$ \\
$d_{m_K}$ (pc)                  & $227\pm 1$ & $230\pm 1$ & $359\pm 2$ & $144\pm 2$ & $231\pm 1$ & $238\pm 1$ \\
$d_{\rm av}^{d}$ (pc)           & $231\pm8$ & $238\pm8$ & $366\pm12$ & $146\pm5$ & $257\pm18$ & $252\pm9$ \\
\enddata
\tablenotetext{a}{The distance is estimated by adopting \teff $=8350$ K, 
\loggl\ $=4.2$ dex, $M_{\star}=2$ \Msuno, and the color excess 
$E(B-V)=0.9$.}

\tablenotetext{b}{The distance is estimated by only changing the 
parameter indicated in the head of each column.}

\tablenotetext{c}{The error bar in the distance is estimated from the
uncertainty in the magnitude.}

\tablenotetext{d}{Average distance weighted by the uncertainty of 
each individual distance determination for the all filters. The error
bar shows the standard deviation from the mean of all the individual
determinations.} 

\end{deluxetable}

\section{Discussion}

\subsection{Membership in IC~348\label{discpm}} 

As shown in Section~\ref{secrot}, the radial velocity of Cernis~52 is
consistent with the mean radial velocity of known members of the
cluster IC~348. 

The membership of a star in a cluster is generally established if the
star is located nearby on the plane of the sky and its radial velocity
and proper motion coincide with those of the cluster.
Fortunately, the proper motion of Cernis~52 has been measured with
high accuracy by R{\"o}ser et al. (2008), providing $(\mu_\alpha \cos
\delta$, $\mu_\delta=(+7.22$,$-8.62)\pm(1.5$,$1.6)$ mas~yr$^{-1}$.

The proper motion of the cluster IC~348 has been determined by Scholz
et al. (1999) in several systems. In particular, using the Hipparcos
system, they derive a mean proper motion of ($\mu_\alpha \cos
\delta$, $\mu_\delta=(+4.6$,$-8.3)\pm(3.7$,$1.6)$ mas~yr$^{-1}$, from nine
Hipparcos stars of the cluster IC~348. 

More recently, the measurement of proper motion of this cluster has
been improved considerably by Loktin \& Beshenov (2003). They
determine a proper motion of $(\mu_\alpha \cos
\delta$, $\mu_\delta=(+6.87$,$-9.15)\pm(0.56$,$0.49)$ mas~yr$^{-1}$.

The radial velocity and proper motion of Cernis~52 strongly support 
its membership in the young cluster IC~348. 
In addition, the nine stars studied by Scholz et al. (1999) provide a
mean distance of $261^{+27}_{-23}$~pc. Therefore, Cernis~52 must be
located at this distance. In the following subsections we will
discuss the mass, radius, age, luminosity and distance of Cernis~52.

\subsection{Mass, Radius and Age\label{discage}}

\begin{figure}[!ht]
\centering
\includegraphics[width=8.cm,angle=0]{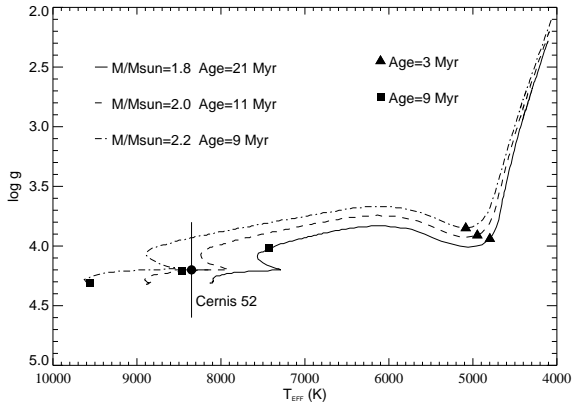}
\caption{Theoretical evolutionary tracks from Siess et al. (2000) for
pre-main sequence stars with masses 1.8, 2.0 and 2.2 \Msun and 
for solar metallicity, $Z=0.02$. The different ages of the tracks
refer to the age of a star at the end of the track, i.e. close the
main sequence. The filled circle represents the 
position of Cernis~52. The triangles and squares following the tracks 
represent the position of the theoretical stars at the age of 3 
and 9 Myr respectively.}    
\label{figtrack}
\end{figure}

In Fig.~\ref{figtrack} we depict theoretical evolutionary tracks for
pre-main sequence stars from Siess et al. (2000). We also overplot the
position of Cernis~52 which seems to be consistent with a theoretical 
$2.0\pm0.2$~\Msun star with an age of $3-20$~Myr. The position of
the star is in between the theoretical tracks with masses 2.2 and 1.8
\Msun. These masses and the derived \logg imply a radius in the range
$1.1<R/R_\odot<2.8$. 

The young cluster IC~348 has an age of 3--7~Myr, according to Trullols
\& Jordi (1997), from a sample of 123 stars in a wide range of
spectral types. Their analysis relies on photometric data and is
possibly affected by unknown extinctions.  
Luhman et al. (2003) discuss the age of IC~348 in detail and
claimed that the observational data seems to imply ages ranging 
from 1 to 10 Myr, although it favours a mean age of 2~Myr. 
They used photometric data of a large sample of K7-M8 spectral-type
stars, although they also obtained spectroscopic data.  
Herbig (1998) also studied the age of roughly 100 stars of this
cluster and found a spread between 0.7 and 12 Myr. All these studies
depend on spectral-type classifications which may be uncertain,
however, they seem to support that the age of IC~348 likely range
between 1 and 12~Myr. 
The position of Cernis~52 in Fig.~\ref{figtrack} is in
agreement with the most likely age range for the cluster IC~348.

\subsection{Luminosity and Distance\label{discdist}}

There is no available paralax for Cernis~52. 
Here, we estimate the distance to Cernis~52 from different 
photometric magnitudes taking into account the stellar parameters.
We know the photometric magnitudes in five
different filters: $m_V=11.4\pm0.1$ (from Cernis
1993), $m_R=10.65\pm0.10$ and $m_I=10.04\pm0.10$ mag, and
$m_J=8.91\pm0.02$, $m_H=8.49\pm0.03$, and $m_{K_{\rm s}}=8.27\pm0.02$
mag from the 2MASS catalog\footnote{http://www.ipac.caltech.edu/2mass/
}. 
Except for the 2MASS magnitudes, we adopted \emph{ad hoc} the values
for the uncertainties on the photometric magnitudes of Cernis~52.  
We derived the radius of the star from the surface gravity, \loggl\
$=4.2\pm0.4$ dex, and assuming a mass of 2~ 
\Msuno. This radius, together with the spectroscopic estimate of the
effective temperature, \teff $= 8350 \pm 200$~K, provides an intrinsic
bolometric luminosity in the range $5.4 < L_\star/L_\odot < 34.2$. 
We determined the absolute bolometric magnitude from the following
formula: $M_{\rm bol,\star}-M_{\rm bol,\odot}=-2.5 \log
(L_\star/L_\odot)$, where the solar bolometric magnitude is
$M_{bol,\odot} = 4.75$ (IAU simposium 1999) and the solar luminosity
is $L_\odot=3.847\,\times\,10^{33}$~erg~s$^{-1}$. 

The apparent bolometric magnitude, $m_{\rm bol,\star}=m_{i,0}-BC_i$,
was derived using the bolometric corrections, $BC_i$, for
non-overshooting ATLAS~9 models (Bessell, Castelli, \& 
Plez 1998). The bolometric corrections were determined as
$BC_i=BC_V+V-i$. For the 2MASS infrared magnitudes we computed the
theoretical colors $V-i$, for $i=J$, $H$, $K_{\rm s}$, from Gonz\'alez
Hern\'andez \& Bonifacio (2009).

We adopt the color excess, $E(B-V)=0.9$ mag, from Cernis 
(1993) which seems to provide consistent distances when using optical
and infrared magnitudes. We compute the magnitude corrected for
extinction in each filter as $m_{i,0}=m_i-A_i$, where $A_i$ were
obtained using the relation $A_i=R_iE(B-V)$, with $R_i$, the ratio of
total to selective extinction, given by the coefficients provided in
McCall (2004). This author gives $R_V=3.07$ with deviations unlikely
to exceed 0.05. Note that using a different value, e.g. $R_V=3.31$ as
derived for Cernis~77 (BD+31$^o$ 643) in Snow et al. (1994) or
$R_V=3.2$ as in Cernis (1993), lead to small corrections to the
distance determination, by --21 and --12~pc respectively. 

The apparent magnitude, $m_i$, was decontaminated from the magnitude
of the companion star, according to the stellar parameters derived in
Section~\ref{seccont}. We estimate the flux ratio of both stars,
$F_{\rm sec}/F_{\rm C52}$, from theoretical magnitudes which were
interpolated in the grid of theoretical colors $V-R$ and $V-I$
(Bessel et al. 1998) and the color $V-K$ and magnitudes $J$, $H$,
$K_{\rm s}$ (Gonz\'alez Hern\'andez \& Bonifacio 2009).
Finally, the corrected magnitude is derived from the following
expression

\begin{displaymath}
m_i=m_{21,i}+\log (F_{21,i}+1)
\end{displaymath}

\noindent
where $m_i$ is the corrected magnitude in the filter $i$, $m_{21,i}$ is
the observed magnitud which contains the flux of both stars, and
$F_{21,i}$, the flux ratio $F_{\rm sec}/F_{\rm C52}$ in the filter
$i$. In Table~\ref{tbldist} we show the derived distance for
each filter with different sets of the relevant parameters.
Note that the relatively small formal error in
the distance estimate for each filter is calculated by
assuming the magnitudes equal $m_i+\Delta m_i$. 

We have derived a distance of 231$^{+135}_{-85}$~pc for the star
Cernis~52, according to the adopted stellar parameters and
mass in section~\ref{seciron} and~\ref{discage}.
Cernis (1993) argued that the star Cernis~52 is a member
of the young cluster IC~348. 
This author derived a distance to Cernis~52 of 236~pc
from the $V$ magnitude, adopting a spectral type of A3V and a color
excess $E(B-V)=0.9$. 
He also derived a distance to the cluster IC~348
of $260\pm16$~pc from 13 probable members of the cluster.
Trullols \& Jordi (1997) provide a photometric determination of the
distance to IC~348 based on the filter $V$ of $240^{+128}_{-84}$~pc
from 43 cluster members. 
These distance determinations rely on spectral type classification. 
Herbig (1998)
argued that spectral types derived from spectroscopy would produce types
somewhat later, and thus smaller $A_V$, and concluded that the 
Trullols \& Jordi (1997) photometric distance should correspond to a
spectroscopic distance of 276~pc. Our distance determination to
Cernis~52 seems to be in agreement with that of the young cluster
IC~348.

To sum up, all the arguments provided in this paper support Cernis~52
as being a member of the young cluster IC~348, and place the star far
enough to explain the large color excess, and the presence of the
interstellar features recently discovered by Iglesias-Groth et
al.~(2008). These intestellar features must be caused by the known
dark clouds already mentioned by Cernis (1993), in particular, the
dark cloud L1470 which covers all the young cluster IC~348. 

\subsection{Photospheric abundances\label{discabu}}

The \ion{O}{1} triplet 6156--8~{\AA} required a high O
abundance ([O/H]\,$=0.5$), whereas the best fit abundance of the \ion{O}{1}
triplet 7771--5~{\AA} is [O/H]\,$=0.15$ (see Fig.~\ref{figo}). This is
unexpected since the \ion{O}{1} triplet at 7771--5~{\AA} shows 
stronger negative NLTE corrections than the \ion{O}{1} triplet
6156--8~{\AA} (Baschek et al. 1977).
This may indicate the existence of some veiling at 7773~{\AA}, instead
of the zero value adopted (according to the extrapolation of the
veiling previously derived as a function of wavelength).
Similar problems are found in the \ion{Mg}{1}b
triplet 5167--83~{\AA} (see Fig.~\ref{figmg}). For these features we 
depict the synthetic spectra computed using the best fit abundance
given by the feature \ion{Mg}{1} 
5173~{\AA}, and this abundance is unable to reproduce both Mg
features at 5167 and 5183~{\AA}. There are other features over the
whole spectrum that cannot be reproduced with the synthetic stellar
spectra which may be related to the presence of unknown DIBs. We
emphasize that for the chemical abundance analysis we have chosen 
only those features that are not affected by any known interstellar
absorption band.

\begin{figure}[!ht]
\centering
\includegraphics[width=7.5cm,angle=0]{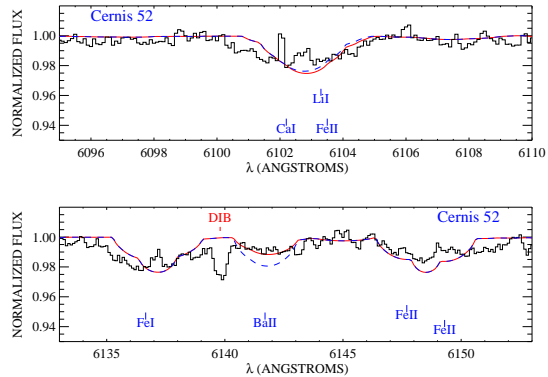}
\caption{Same as Fig.~\ref{figmg}, but for other spectral
regions. Upper panel: synthetic spectra are computed with
A(Li)\,$=4.0$ (solid line) and A(Li)\,$=3.3$ (dashed line). We
emphasize that A(Li)\,$=3.3$ must be the maximum Li abundance in this
star (see Section~\ref{disnaph}). 
Bottom panel: synthetic spectra are computed with
[Ba/H]\,$=-0.8$ (solid line) and [Ba/H]\,$=-0.1$ (dashed line). 
}
\label{figba}
\end{figure}

The photospheric abundances of the star seems to be almost solar
within the error bars, i.e., the star does not belong to the group of
chemically peculiar (CP) A-type stars. 
There appears to be a correlation between the presence of chemical
anomalies and the rotational velocity of these stars (e.g. Fossati et al.
2008; Takeda et al. 2008). Thus, stars with $v\sin i \lesssim
50-75$~\kms are underabundant in O and Ca (down to [X/H]$\sim-0.8$),
and overabundant in Fe ([Fe/H]$\sim0.5$) and specially Ba
(up to [Ba/H]$\sim1.5$), whereas Si and S shows near-solar abundances.
Cernis~52 has a rotational velocity on the edge between ``peculiar'' and
``normal'' A stars. In the bottom panel of Fig.~\ref{figba}, we depict
the Ba line which gives an abundance [Ba/H]$\sim -0.8$, which is even
lower than the solar abundance. This definitively discards Cernis~52
as a chemically peculiar A-type star.

\begin{figure}[!ht]
\centering
\includegraphics[width=7.5cm,angle=0]{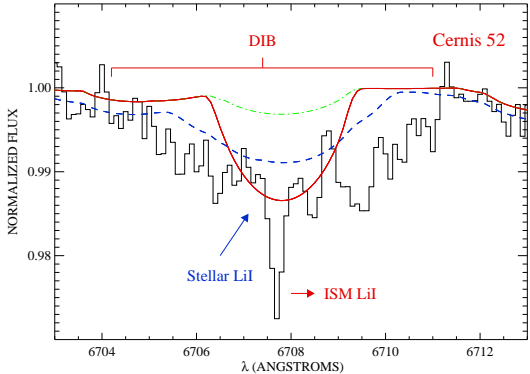}
\caption{Same as Fig.~\ref{figmg}, but for other spectral
region. Synthetic spectra are computed with A(Li)\,$=4.0$
(solid line) and A(Li)\,$=3.3$ (dashed-dotted line). 
We also display the combined spectrum (dashed line) including
the spectrum of Cernis~52 contributing with 90\% of the
stellar flux and the spectrum of the companion star contributing at
10\%, respectively. The dashed line shows a synthetic spectrum with
A(Li)\,$=3.3$ and 4.0, for Cernis~52 and the companion, respectively.
}
\label{figli}
\end{figure}

\subsection{The naphthalene feature at 6707.4~{\AA}\label{disnaph}}

In our paper announcing the discovery of the naphthalene cation 
(Iglesias-Groth et al. 2008), we noted that the strongest
band of the interstellar cation appeared at 6707.4 {\AA}, i.e. very
close to the stellar \ion{Li}{1} resonance line.   
The abundance analysis presented in this paper was expected to 
provide indications of the likely Li abundance for the star, i.e., a
star without abundance anomalies is very likely to have a lithium
abundance no greater than the local abundance of A(Li)$ =3.3$ (see
bottom panel of Fig.~\ref{figsi}).

In Fig.~\ref{figli} we show three synthetic
spectra: (i) two of them computed only considering the star Cernis~52
and with a already too high Li abundance, A(Li)\,$=4.0$ and a cosmic
Li abundance, A(Li)\,$=3.3$, and, (ii) the
other combining the spectrum of Cernis~52 with A(Li)\,$=3.3$ and 
the spectrum of the companion star with A(Li)\,$=4.0$, contributing
with 90 and 10\%, respectively, to the stellar flux.
We estimated the rotational velocity (see Section~\ref{secrot}) at
$v\sin i \sim 65$~\kms from many stellar lines in the spectral range
$\lambda\lambda5270-6400$\,{\AA}. As seen in Fig.~\ref{figsi} and more
clearly in Fig.~\ref{figli}, the stellar lithium line at 6707.8 {\AA}
is unable to fill the whole feature even if we increase the Li
abundance up to A(Li)\,$=4.0$.  
In the second case, we choose a rotational velocity for the
companion star of 100~\kms to reproduce this broad feature, but even
using a very high Li abundance, A(Li)\,$=4.0$, is not enough to
completely fill this broad feature.  The required Li abundance able to
fit this feature, A(Li)\,$=5.6$, is extraordinarily high.  
Again A(Li)\,$=3.3$ is a highly probably maximum for the companion at
\teff$=5800$\,K.
We must emphasize that even with this large rotational velocity and
abnormally high Li abundance for the companion star, the feature is
still not perfectly reproduced by the synthetic spectrum. 

In the upper panel of Fig.~\ref{figba}, we depict the subordinate
\ion{Li}{1} line at 6103.4~{\AA}. The subordinate Li line is very weak
and the feature at 6103~{\AA} is dominated by a \ion{Ca}{1} line at
6102.2~{\AA} and a \ion{Fe}{2} line at 6103.5~{\AA}. 
Thus, increasing the Li abundance from
A(Li)\,$=3.3$ to A(Li)\,$=4.0$ has a negligible effect on the line
profile. Thus, we cannot estimate the Li abundance from this
subordinate Li line. Polosukhina \& Shavrina (2007) suggested that the
Li abundance obtained in CP stars with magnetic fields from the
Li subordinate line may be higher than that derived from the 
resonance Li line by 0.2--0.4 dex. 
This may suggest vertical stratification of Li. In addition, CP
stars shows enhanced Li abudances up to A(Li)\,$=4.0$. 

In Section~\ref{discabu} we discarded the possibility that Cernis~52
is a CP star, so its Li abundance must be at the most the cosmic Li
abundance, i.e. A(Li)\,$=3.3$. Therefore, as seen in Fig.~\ref{figli},
the stellar Li line should only be responsible for a small fraction of
the equivalent width of the broad feature at 6707.4~{\AA}.

All the above statements confirm that this broad feature must have a
insterstellar origin and most probably linked to the naphthalene
molecule. The other broad feature at 6756~{\AA} (see Fig.~\ref{figsi})
is probably another interstellar feature (S. Iglesias-Groth et al. 2009,
in preparation). 

Finally, another detection of PAHs in the interstellar medium has
been recently discovered by Iglesias-Groth et al. (2009).
These authors report a PAH band at $\sim 7088$~{\AA}, the
anthracene cation, ${\rm C}_{14}{\rm H}_{10}^+$. 

\section{Conclusions}
 
We have performed a detailed chemical analysis of the star Cernis~52.
We apply a technique that provides a determination of the stellar
parameters, taking into account any possible source of veiling. 
We find $T_{\mathrm{eff}} = 8350 \pm 200$ K, $\log
(g/{\rm cm~s}^2) = 4.2 \pm 0.4$, $\mathrm{[Fe/H]} = -0.01 \pm 0.15$, 
and a veiling (defined as $F_{\rm  veil}/F_{\rm total}$) of less
than 55\% at 5000 {\AA} and decreasing toward longer wavelengths. 

The spectrum of Cernis~52 shows many features very likely related with
the interstellar medium. In addition, we discover in photometric
images a companion 1.7~mag fainter star at a distance of
$0.818\arcsec\pm0.007\arcsec$, but this star only contributes with
10\% of the stellar flux in our spectra and hardly affect the stellar
features of Cernis~52. The derived chemical abundances are
roughly solar within their error bars. This prevent the star from
being a chemically peculiar star. 

We have determined the radial velocity of Cernis~52 at $v_r=+13.7 \pm 
1$~\kmso, being almost equal to the mean radial velocity of the
young cluster IC~348. 

We have compared the stellar parameters with pre-main-sequence
evolutionary tracks of solar metallicity and see that the star is
consistent with being a pre-main-sequence A-type star with an age of
$3-20$~Myr.  

We have estimated the distance to Cernis~52 using the available
Johnson-Cousins and 2MASS photometric data, at
231$^{+135}_{-85}$~pc according to the stellar parameters and its
error bars. This value also agrees with the distance to the cluster
IC~348. 

The proper motion of Cernis~52, $(\mu_\alpha \cos
\delta$, $\mu_\delta=(+7.22$,$-8.62)\pm(1.5$,$1.6)$ mas~yr$^{-1}$, is
consistent with the proper motion of IC~348.

All these measurements make it likely that the star Cernis~52
(\mbox{BD+31$^o$ 640}) belongs to the young cluster IC~348. 

We confirm that the feature at 6707.4~{\AA} is not related with
a stellar lithium line because the rotational velocity of the star,
$v\sin i=65 \pm 5$~\kmso, is too low to explain the broad feature
associated with the naphthalene cation. Furthermore, the presence of
a companion star cannot either explain this feature even for an
abnormally high Li abundance, $\log [N({\rm Li})/N({\rm H})]+12 > 5$
dex.

As already stated in Cernis (1993), the interstellar features, in
particular, the naphthalene cation, that appear in the spectrum of
Cernis~52 may form in the dark cloud L1470 which covers all the
cluster IC~348 and is at about the same distance.

\acknowledgments

J.I.G.H. acknowledges support from the EU contract MEXT-CT-2004-014265
(CIFIST). D.L.L. thanks the Robert A. Welch Foundation of
Houston, Texas, for support through grant F-634. D.A.G.H. acknowledges
support from the Spanish Ministry of Science and Innovation (MICINN)
under the 2008 Juan de la Cierva Programme.
This work has been partially founded by project
AYA2007-64748 of the Spanish Ministry of Education and Science. 
We thank the FastCam team at Instituto de Astrof{\'\i}sica
de Canarias and Universidad Polit\'ecnica de Cartagena for kindly
obtaining images of Cernis 52 for this work.  
We are grateful to V{\'\i}ctor J. S. B\'ejar and Jos\'e A.
Caballero for helpful discussions. 
We are grateful to Tom Marsh for the use of the MOLLY analysis
package. This work has made use of the VALD, SIMBAD, DSS and 2MASS
databases, and IRAF and Aladin facilities. 
The Two Micron All Sky Survey is a joint project of
the University of Massachusetts and the Infrared Processing and
Analysis Center/California Institute of Technology, funded by the
National Aeronautics and Space Administration (NASA) and the National
Science Foundation (NSF). 
The compressed files of the "Palomar Observatory -
Space Telescope Science Institute Digital Sky Survey" of the northern
sky, based on scans of the Second Palomar Sky Survey are copyright (c)
1993-2000 by the California Institute of Technology and are
distributed herein by agreement. All Rights Reserved.
Produced under Contract No. NAS5-2555 with the National Aeronautics
and Space Administration.

\end{document}